\documentclass[aps,prd,
twocolumn,
groupedaddress,
nofootinbib,
amsmath,amssymb]{revtex4-2}

\usepackage[utf8]{inputenc}

\usepackage{graphicx}
\usepackage[colorlinks=true, allcolors=blue]{hyperref}
\usepackage{physics}
\usepackage{mathtools}
\usepackage{siunitx}
\usepackage{bm}
\usepackage{comment}
\usepackage{booktabs}
\usepackage{tikz}
\usetikzlibrary{arrows.meta}
\newcommand{\identity}{\text{\usefont{U}{bbold}{m}{n}1}}
\MakeRobust{\identity}
\mathtoolsset{showonlyrefs}
\usepackage{lipsum}

\usepackage{xcolor}
\definecolor{my0}{HTML}{226F99}
\definecolor{my1}{HTML}{F77A3B}
\definecolor{my2}{HTML}{55AD6E}
\definecolor{my3}{HTML}{C795C6}
\definecolor{my4}{HTML}{EBCC36}
\definecolor{mygrey}{HTML}{999999}

\def\[#1\]{\begin{equation}\begin{aligned}[b]#1\end{aligned}\end{equation}} 

\newcommand{\dirac}{\delta_\mathrm{D}}
\newcommand{\Om}{\ensuremath\Omega_\mathrm{m}}

\newcommand{\Ol}{\ensuremath\Omega_\Lambda}
\newcommand{\transp}{\top}

\newcommand{\ini}{{(\mathrm{i})}}
\newcommand{\fin}{{(\mathrm{f})}}
\newcommand{\free}{{(0)}}
\newcommand{\xcl}{\bm{x}_\mathrm{cl}}
\newcommand{\Z}{\mathcal{Z}}
\newcommand{\DD}{\mathcal{D}}
\newcommand{\laplace}{\Delta}
\newcommand{\vPi}{\vec{\Pi}}
\newcommand{\im}{\mathrm{i}}
\newcommand{\e}{\mathrm{e}}
\newcommand{\I}{\mathrm{I}}
\newcommand{\B}{\mathcal{B}}
\newcommand{\V}{\mathrm{v}}

\newcommand{\conn}{\mathrm{c}}
\renewcommand{\H}{\mathcal{H}}
\newcommand{\mat}[1]{\mathrm{#1}}
\newcommand{\gqp}{g_{qp}}
\newcommand{\gpp}{g_{pp}}

\begin{document}
\title{Cosmic structure from the path integral of classical mechanics\\ and its comparison to standard perturbation theory}
\author{Marvin Sipp}
\email[]{sipp@thphys.uni-heidelberg.de}
\author{Hannes Heisler}
\email[]{heisler@thphys.uni-heidelberg.de}
\author{Matthias Bartelmann}
\affiliation{Institut für Theoretische Physik, Universität Heidelberg, Philosophenweg 12, 69120 Heidelberg, Germany}

\date{9 February 2026}

\makeatletter
\hypersetup{ 
	pdfauthor={Marvin Sipp},
	pdftitle=\@title
}
\makeatother

\begin{abstract}
We investigate cosmic structure formation in the framework of a path-integral formulation of an $N$-particle ensemble in phase space, dubbed resummed kinetic field theory (RKFT), up to one-loop perturbative order. In particular, we compute power spectra of the density contrast, the divergence and curl of the momentum density and arbitrary $n$-point cumulants of the stress tensor. In contrast to earlier works, we propose a different method of sampling initial conditions, with a Gaussian initial phase-space density. Doing so, we exactly reproduce the corresponding results from Eulerian standard perturbation theory (SPT) at one-loop order, showing that formerly found deviations can be fully attributed to inconsistencies in the previous sampling method.
Since, in contrast to SPT, the full phase-space description does not assume a truncation of the Vlasov hierarchy, our findings suggest that nonperturbative techniques are required to accurately capture the physics of cosmic structure formation. 
\end{abstract}

\maketitle


\section{Introduction}
According to the concordance model of cosmology~\cite{Bartelmann2010}, the majority of the Universe's matter content does not partake in the electromagnetic interaction and is therefore called \emph{dark matter}. However, the fundamental nature of dark matter is not known. Under the influence of gravity, initially small fluctuations in the otherwise homogeneous and isotropic matter distribution are enhanced to form the cosmic large-scale structure observed today. Therefore, the latter is one of the most important observables in cosmology, having information on both dark matter and the laws of gravity on large scales imprinted on their statistics.
Cosmic structure formation, however, is a highly nonlinear process. Numerical simulations are successfully being employed in this context~\cite[e.\,g.][]{Springel2005,Klypin2011,Springel2017,Wang2020}. Despite their tremendous success, their computational cost makes it unfeasible to efficiently scan through different cosmological models and theories of dark matter and gravity. In order to overcome this shortcoming and to understand the physical processes at play, an analytical approach is desirable.
The most prominent method, Eulerian standard perturbation theory (SPT), is based upon an Euler-Poisson system of equations for an ideal dark matter fluid. A perturbative expansion is constructed in the density contrast and peculiar velocity field~(see e.\,g.~\cite{Bernardeau2002} for a review). This perturbative expansion, however, breaks down at small scales, where the density contrast is of order unity. Furthermore, the ideal fluid ansatz cannot describe the effects of stream crossing. Going beyond the single-stream approximation requires a phase-space description.
The full dynamics of the dark matter phase-space distribution are described by the nonlinear Vlasov-Poisson system~\cite{Peebles1980}. The single-stream approximation of SPT is equivalent to a truncation of the moment hierarchy of the Vlasov equation at the first nontrivial order. There have been various efforts to include higher moments, either by a different truncation scheme~\cite{Erschfeld2019,Garny2025}, by directly perturbing the Vlasov equation~\cite{Valageas2001,Nascimento2025}, or by absorbing the effects of higher-order moments into operators of an effective field theory (EFT)~\cite{Baumann2012,Carrasco2012}. Furthermore, a variety of nonperturbative methods have been applied in this context~\cite{Crocce2006,Pietroni2008,Floerchinger2017,Erschfeld2024}.
A complementary approach, Lagrangian perturbation theory (LPT), is based on the Lagrangian instead of the Eulerian description of dark matter. The perturbative expansion is then made in the displacement field relative to the initial positions of the fluid elements or particles~\cite{Buchert1992,Buchert1993,Buchert1994,Porto2014,Rampf2021}. However, the mapping of the initial to the evolved density contrast by means of mass conservation becomes singular when streams cross.

In recent years, a new formalism has been developed, which is in spirit much closer to $N$-body simulations: Based on the path-integral formulation of classical Hamiltonian mechanics developed in~\cite{Gozzi1988,Gozzi1989,Das2012,Das2015}, the dark matter distribution is modeled as an $N$-particle ensemble in phase space~\cite{Fabis2015,Bartelmann2016,Fabis2018,Bartelmann2019}. This approach has been dubbed kinetic field theory (KFT). 
In~\cite{Lilow2018,Lilow2019}, an exact reformulation of the theory in terms of the associated phase-space density instead of individual particle trajectories was developed, resummed KFT (RKFT). The resulting density power spectra up to one-loop perturbative order differ from those of SPT by a small-scale damping~\cite{Lilow2018,Lilow2019,Daus2025}. Since perfectly cold initial conditions were assumed, this is somewhat surprising. In the large-$N$ limit, the underlying dynamics should be equivalent to the Vlasov equation~\cite{Chavanis2008}. As the single-stream approximation poses a ``fixed point'' of the momentum-cumulant hierarchy of the Vlasov equation, perturbative solutions with perfectly cold initial conditions should recover SPT~\cite{Pueblas2009}.

In this work, we show that these differences can be fully traced back to how the initial conditions of the particle ensemble were sampled. The paper is structured as follows: In Sec.~\ref{sec:theory}, we recapitulate the formalism of (R)KFT, and review the original derivation of the KFT initial conditions in Sec.~\ref{sec:old_ICs}, pointing out an inconsistent approximation in the procedure of choosing initial momenta. Thereafter, we introduce a different sampling method, with a Gaussian initial phase-space density, in Sec.~\ref{sec:new_ICs}. We compare tree-level density and momentum spectra for both sets of initial conditions in Sec.~\ref{sec:tree_level}, where we show that they differ by the aforementioned small-scale damping. In fact, for the new initial conditions, the results agree exactly with linear SPT, as expected. In Sec.~\ref{sec:one_loop}, we show that this equivalence extends to one-loop order and also to arbitrary one-loop $n$-point cumulants of the stress tensor. Finally, we discuss the implications of this work and give an outlook to future studies in Sec.~\ref{sec:conclusion}.

\section{Theoretical Framework}\label{sec:theory}
For completeness, and for readers unfamiliar with (R)KFT, we summarize the theoretical framework developed in~\cite{Fabis2015,Bartelmann2016,Lilow2018,Lilow2019} in this section. See also~\cite{Bartelmann2019} for a comprehensive review.

\subsection{Basic setup and notation}
Consider a system of $N$ classical particles of mass~$m$, each labeled by an index $j=1,\dots,N$. The state of each particle is described by its phase-space coordinates~$x_j = (\vec{q}_j,\vec{p}_j)^\transp$, with canonical positions $\vec{q}_j$ and momenta $\vec{p}_j$, in general depending on time $t$. Let now the Hamiltonian of the $N$-particle system be of the form
\[\label{eq:hamiltonian}
    H(x_1,\dots,x_N,t) &= H_0(x_1,\dots,x_N) + V(x_1,\dots,x_N,t) \\
    &= \sum_{j=1}^{N} \frac{p_j^2}{2m} + V(x_1,\dots,x_N,t).
\]
In absence of an external potential, the interaction shall be given in terms of a two-particle potential $v$ that depends only on the distance between the particles and potentially time,
\[
    V(x_1,\dots,x_N,t) = \frac{1}{2} \sum_{i\neq j=1}^{N} v(\abs{\vec{q}_i-\vec{q}_j},t).
\]
Each particle, $j=1,\dots,N$, is governed by its Hamiltonian equations of motion,
\[
    \dot{x}_j = \omega\nabla_j H(x_1,\dots,x_N,t) \qq{with} \omega = \begin{pmatrix}
        0 & \identity_3 \\ -\identity_3 & 0
    \end{pmatrix},
\]
where a dot denotes the derivative with respect to time and ${\nabla_j = \qty(\nabla_{\vec{q}_j}, \nabla_{\vec{p}_j})^\transp}$.
The state of the full $N$-particle system is then described by the $6N$-dimensional bundle of phase-space trajectories, which we compactly write as
\[\label{eq:bundle_notation}
    \bm{x} = \sum_{j=1}^{N} x_j \otimes e_j,
\]
where $\{e_j\}$ is the $N$-dimensional canonical unit basis. The tensor product induces a standard inner product. This boldface notation straightforwardly generalizes to all vectors, scalars, integral measures and derivative operators. Subscripts $\bm{q}$ and $\bm{p}$ shall indicate a restriction to the position and momentum subspaces, respectively, and $\bm{q} \equiv \bm{x}_{\bm{q}}$, $\bm{p} \equiv \bm{x}_{\bm{p}}$. The equations of motion of the system then read
\[\label{eq:bundle_eom}
    \dot{\bm{x}} - \bm{\omega}\bm{\nabla}H(\bm{x},t) = 0,
\]
with the $N$-particle generalization of the symplectic matrix, $\bm{\omega} = \omega\otimes\identity_N$. Given some initial state~$\bm{x}^\ini$ at time~$t^\ini$, the system can be evolved to later times by solving the equations of motion \eqref{eq:bundle_eom}. In practice, however, the exact initial conditions of a system consisting of a large number of particles are almost never known. Instead, they are typically characterized by some initial phase-space density $\varrho_N(\bm{x}^\ini)$. The phase-space density at a later time is then obtained by evolving all particles along their classical trajectories,
\[\label{eq:psd_evol}
    \varrho_N(\bm{x},t) = \int\dd\bm{x}^\ini\,\dirac\qty(\bm{x}-\xcl(t; \bm{x}^\ini))\, \varrho_N(\bm{x}^\ini),
\] 
where the Dirac distribution enforces the solution~$\xcl(t; \bm{x}^\ini)$ of the Hamiltonian equations of motion~\eqref{eq:bundle_eom}, starting from the initial conditions $\bm{x}^\ini$. It can be thought of as the deterministic classical equivalent to the transition probability of going from an initial state~$\bm{x}^\ini$ to some final state at a later time.

Consider now some observable 
$\mathcal{O}(x,t) \equiv \mathcal{O}(x;\bm{x}(t))$,
depending parametrically on the time-dependent phase-space configuration of the system and possibly on phase-space position $x$. Its expectation value can be calculated as ensemble average,
\[\label{eq:ev_observables}
    \ev{\mathcal{O}(x,t)} &= \int\dd\bm{x}\,\mathcal{O}(x;\bm{x}) \varrho_N(\bm{x},t) \\
    &\overset{\eqref{eq:psd_evol}}{=} \int\dd\bm{x}^\ini\,\mathcal{O}(x;\xcl(t; \bm{x}^\ini))\,\varrho_N(\bm{x}^\ini).
\]
This straightforwardly generalizes to arbitrary $n$-point functions.
In the following, we will omit the parametric argument on observables whenever there is no danger of confusion.
An important example for an ``observable'' of interest is the Klimontovich phase-space density
\[\label{eq:klimontovich}
    f(x,t) = \sum_{j=1}^{N} \dirac(x - x_j(t)),
\]
where $x = (\vec{q},\vec{p})^\transp$. From this we can take momentum moments to get e.\,g.~the particle number density $\rho$, momentum density $\vPi$, and the stress tensor $\mat{T}$,
\[\label{eq:momentum_moments}
    \rho(\vec{q},t) &= \int \dd[3]p\,f(\vec{q},\vec{p},t)\\
    \vPi(\vec{q},t) &= \int \dd[3]p\,\vec{p}\, f(\vec{q},\vec{p},t)\\ 
    \mat{T}(\vec{q},t) &= \int \dd[3]p\,(\vec{p}\otimes\vec{p})\, f(\vec{q},\vec{p},t)
\]
In homogeneous systems, we are often interested in the statistics of fluctuations around a mean value. For the density, for example, we have
\[
    \ev{\rho\rho} = \ev{(1+\delta)\bar{\rho}\,(1+\delta)\bar{\rho}} = \bar{\rho}^2 (1+ \ev{\delta\delta}),
\]
where $\bar{\rho} = \ev{\rho}$ and $\delta=\rho/\bar{\rho}-1$ denotes the density contrast with $\ev{\delta} = 0$. The autocorrelation of the density contrast is proportional to the connected two-point function of the density,
\[
    \bar{\rho}^2\ev{\delta\delta} = \ev{\rho\rho} - \ev{\rho}\!\ev{\rho} = \ev{\rho\rho}_\conn.
\]
We denote general connected correlation functions (i.\,e.~cumulants) with
\[
    G_{\mathcal{O}_1\dots\mathcal{O}_n} \equiv \ev{\mathcal{O}_1\dots\mathcal{O}_n}_\conn.
\]
Computing the $n$-point statistics of interest from~\eqref{eq:ev_observables} would require the full solution to the equations of motion for the $N$-particle system, which, in practice, is almost never obtainable. However, the problem can be reformulated in terms of the path-integral of classical mechanics, motivating the use of perturbative and nonperturbative techniques known from quantum field theory (QFT) and condensed matter physics.

\subsection{Path-integral approach to kinetic theory}\label{sec:micro_theory}
As usual in statistical mechanics, we define the partition function $\Z$ as the normalization of the density of states, in our case the phase-space density,
\[
    \Z = \int\dd\bm{x}^\fin \varrho_N(\bm{x}^\fin,t^\fin),
\]
where $\bm{x}^\fin$ denotes the phase-space configuration at some ``final'' time $t^\fin$.
Discretizing time on a regular grid between $t^\fin$ and $t^\ini$, at which $\varrho_N(\bm{x}^\ini)$ is defined, and plugging in the time evolution~\eqref{eq:psd_evol}, one can define a path integral as the continuum limit of infinitesimal time steps, in complete analogy to the procedure known from the path integral formulation of quantum mechanics. The partition function then reads
\[\label{eq:Z_micro}
    \Z &= \int\dd\bm{x}^\fin \,\dd\Gamma^\ini \int_{\bm{x}^\ini}^{\bm{x}^\fin} \DD\bm{x}\, \dirac\qty[\bm{x}(t)-\xcl(t; \bm{x}^\ini)]\\
    &= \int\dd\bm{x}^\fin \,\dd\Gamma^\ini \int_{\bm{x}^\ini}^{\bm{x}^\fin} \DD\bm{x}\, \dirac\qty[\dot{\bm{x}}(t) - \bm{\omega}\bm{\nabla}H(\bm{x},t)] \\
    &= \int\dd\bm{x}^\fin \,\dd\Gamma^\ini \int_{\bm{x}^\ini}^{\bm{x}^\fin} \DD\bm{x}\,\DD\bm{\chi}\, \exp(\im S[\bm{x,\bm{\chi}}]).
\]
with $\dd\Gamma^\ini \equiv \dd\bm{x}^\ini\varrho_N(\bm{x}^\ini)$, and
where the path integral is over all possible phase-space trajectories of the $N$-particle ensemble with fixed initial and final configurations.
The functional Dirac distribution appearing in the first line singles out those trajectories which solve the Hamiltonian equations of motion of the system.
In practice, however, we will not be able to solve the full trajectories of all particles. Instead, we replaced the argument of the Dirac distribution with the equations of motion in the second step\footnote{Here, we would usually get a functional determinant, but it has been shown that it can be set to unity in our case. This choice actually requires us to define $\Theta(0) = 0$ for the Heaviside function. See \cite{ZinnJustin1986,Gozzi2000} for more detail.}.
Finally, to further strengthen the correspondence with QFT, we expressed the Dirac distribution in terms of an auxiliary field $\bm{\chi}$ conjugate to the equations of motion~\cite{Martin1973}, defining the \emph{action}
\[\label{eq:action_full}
    S[\bm{x},\bm{\chi}] = \int_{t^\ini}^{t^\fin} \dd t\, \bigg[\bm{\chi}(t)\cdot\big(\dot{\bm{x}}(t) - \bm{\omega}\bm{\nabla}H(\bm{x},t)\big)\bigg].
\]
The $n$-point statistics of the system can now be generated by either introducing sources for $\bm{x}$ and $\bm{\chi}$ and acting with suitable functional derivatives on the so constructed generating functional, or by directly inserting the observables into the integrand of~\eqref{eq:Z_micro}.
Indeed, doing so and solving the path integral, we recover~\eqref{eq:ev_observables}.
In general, the evolution of a self-interacting system will be highly nonlinear such that it cannot be solved exactly. The free evolution, however, is a rather simple problem. Let us therefore insert the Hamiltonian~\eqref{eq:hamiltonian} into the action~\eqref{eq:action_full} in order to split it into a free and an interacting part,
\[
    S[\bm{x},\bm{\chi}] &= S_0[\bm{x},\bm{\chi}] + S_\I[\bm{x},\bm{\chi}]\\
    S_0[\bm{x},\bm{\chi}] &= \int_{t^\ini}^{t^\fin} \dd t\, \bigg[\bm{\chi}(t)\cdot\big(\dot{\bm{x}}(t) - \bm{\omega}\bm{\nabla} H_0(\bm{x},t)\big)\bigg]\\
    S_\I[\bm{x},\bm{\chi}] &= \int_{t^\ini}^{t^\fin} \dd t\, \bigg[\bm{\chi}_{\bm{p}}(t)\cdot\bm{\nabla}_{\bm{q}} V(\bm{q},t)\bigg].
\]
The partition function can then be expanded in powers of the interaction,
\begin{multline}
    \label{eq:micro_pt}
    \Z = \int\dd\bm{x}^\fin \,\dd\Gamma^\ini \int_{\bm{x}^\ini}^{\bm{x}^\fin} \DD\bm{x}\,\DD\bm{\chi}\, \exp(\im S_0[\bm{x,\bm{\chi}}]) \\
    \times \sum_{k=0}^{\infty} \frac{\im^k}{k!} S_\I^k[\bm{x},\bm{\chi}],
\end{multline}
where the first line defines the free partition function~$\Z^\free$. The free theory is easily solved, since in absence of forces, the particle trajectories are simply straight lines, i.\,e.~$\xcl(t) \to \bm{x}^\free(t)$ with components
\[\label{eq:free_trajectories}
    \bm{q}^\free(t; \bm{x}^\ini) &= \bm{q}^\ini + t\,\frac{\bm{p}^\ini}{m} \qq{and} \bm{p}^\free(t; \bm{x}^\ini) = \bm{p}^\ini.
\]
Solving the free theory reduces to plugging this into~\eqref{eq:ev_observables}, yielding expectation values of observables for an ensemble of freely streaming particles.
A perturbative scheme in powers of the interaction potential is then established by truncating the sum in the second line of~\eqref{eq:micro_pt} at higher orders.

Indeed, \eqref{eq:micro_pt} contains the full phase-space dynamics of the interacting system, equivalently to the Liouville equation.
We want to stress that this is fundamentally different to SPT, where the perturbation theory is in powers of the density contrast and velocity fluctuations. The underlying equations of motion assume a fixed truncation of the Vlasov hierarchy from the onset. Thus, the full phase-space information is never available in SPT.
There are, however, also some disadvantages to the KFT perturbation theory described above. First of all, the theory is not formulated in terms of the \emph{macroscopic} fields, i.\,e.~$f$, $\rho$ or $\vPi$, whose statistics we are interested in, but in terms of their constituent particles' trajectories. For this reason, the perturbative expansion scheme derived from~\eqref{eq:micro_pt} was dubbed \emph{microscopic perturbation theory} ($\mu$PT). This is an oddity, but not an issue \emph{per se}, since the macroscopic fields are (generalized) functions with parametric dependence on the particle trajectories.
The more important issue, however, is that when applied to cosmic structure formation, $\mu$PT converges only very slowly to numerical results \cite{Pixius2022,Heisenberg2022}. It turns out that these two points can be addressed simultaneously.

\subsection{Macroscopic reformulation}\label{sec:macro_pt}
In this subsection, we summarize the key steps in the derivation of RKFT, which was first developed in~\cite{Lilow2018,Lilow2019}.
As explained in the previous section, the free theory is solved exactly without much effort. Attention shall now be directed toward the interacting part of the action. It can indeed be reformulated in terms of macroscopic fields~\cite{Fabis2015,Bartelmann2016},
\[\label{eq:S_I_doublet}
    \im S_\I[\bm{x},\bm{\chi}] &= \im \int_{t^\ini}^{t^\fin} \dd t\, \bigg[\bm{\chi}_{\bm{p}}(t)\cdot\bm{\nabla}_{\bm{q}} V(\bm{q},t)\bigg]\\
    &= \int \dd X f(X)\B(X) \equiv f\cdot\B\\
\]
consisting of the Klimontovich phase space density $f$ defined in~\eqref{eq:klimontovich} and the \emph{response field}
\[\label{eq:response_field}
    \B(\vec{q},\vec{p},t) \coloneq \im \sum_{j=1}^N \vec{\chi}_{\vec{p}_j}(t)\cdot\nabla_{\vec{q}_j} v(\abs{\vec{q}-\vec{q}_j}, t),
\]
describing the forces acting on a given point in phase space. Furthermore, we adopted a collective notation for extended phase-space coordinates, $X = (x,t)^\transp = (\vec{q},\vec{p},t)^\transp$.

The theory can be ``coarse-grained'' as explained in Appendix~\ref{app:rkft_derivation}, resulting in an exact reformulation of the theory in terms of the macroscopic phase-space density~$\psi_f$ and response field~$\psi_\B$, which obey the same statistics as $f$ and $\B$, respectively~\cite{Lilow2018,Lilow2019,Daus2024}. Its generating functional reads
\[
    \Z[J] = \int\DD\psi \exp(-\frac{1}{2}\psi^\transp\,\Delta^{-1}\,\psi + S_\V[\psi] + J\cdot\psi),
\]
with $\psi = (\psi_f,\psi_\B)^\transp$ and the source doublet $J$.
The microscopic information is contained in the propagator~$\Delta$ and the vertex part of the action~$S_\mathrm{v}$ in terms of the cumulants of the free theory,
\[
    G_{f\dots f\B\dots\B}^\free = \ev{f\dots f\B\dots\B}_{\mathrm{c},0},
\]
derived via the free partition function~$\Z^\free$ defined in the previous section.
The inverse propagator is given by
\[
    \Delta^{-1} &= \begin{pmatrix}
        0 & \identity - G_{\B f}^\free\\
        \identity - G_{f\B}^\free & -G_{ff}^\free
    \end{pmatrix}.
\]
Inverting it yields
\[
    \Delta &= \begin{pmatrix}
        \Delta_{ff} & \Delta_{f\B}\\
        \Delta_{\B f} & 0
    \end{pmatrix}
    \equiv \begin{pmatrix}
        \begin{tikzpicture}
            \draw[thick, {Latex[length=3mm/2,width=3mm/2]}-{Latex[length=3mm/2,width=3mm/2]}] (.25/2,0) -- (1.75/2,0);
            \draw[thick] (0,0) -- (1,0);
            \filldraw (.5,0) circle (.05);
        \end{tikzpicture}
        &
        \begin{tikzpicture}
            \draw[thick, {Latex[length=3mm/2,width=3mm/2]}-] (.75/2,0) -- (1,0);
            \draw[thick] (0,0) -- (1,0);
        \end{tikzpicture}
        \\
        \begin{tikzpicture}
            \draw[thick, -{Latex[length=3mm/2,width=3mm/2]}] (0,0) -- (1.25/2,0);
            \draw[thick] (0,0) -- (1,0);
        \end{tikzpicture}
        &
        0
    \end{pmatrix}
    ,
\]
with the components
\[\label{eq:def_resummed_propagator}
    \Delta_{ff}(X_1,X_2) &= \qty(\Delta_{f\B}\,G_{ff}^\free\,\Delta_{\B f})(X_1,X_2)\\
    \Delta_{f\B}(X_1,X_2) &= \Delta_{\B f}(X_2,X_1) = \qty(\identity-G_{f\B}^\free)^{-1}(X_1,X_2).
\]
The remaining inversion is in the functional sense. Formally, it is solved by a Neumann series,
\[\label{eq:Neumann_series}
    \Delta_{f\B}(X_1,X_2) &= \sum_{n=0}^{\infty} \qty[G_{f\B}^\free]^n (X_1,X_2) \\
    &\equiv \identity(X_1,X_2) + \widetilde\Delta_{f\B}(X_1,X_2).
\]
In the $\mu$PT expansion~\eqref{eq:micro_pt}, the perturbative order of any term corresponds to the number of $\B$-fields appearing in it.
Thus, already at the level of the propagator, the reformulated theory contains a partial resummation of the $\mu$PT series. For this reason, it was dubbed \emph{resummed KFT} (RKFT) in the original paper \cite{Lilow2019}.
In practice, $\Delta_{f\B}$ is computed by reinserting this ansatz into~\eqref{eq:def_resummed_propagator}, obtaining a self-consistency equation for $\widetilde{\Delta}_{f\B}$. We show this in Appendix~\ref{app:macro_details}.
The vertices consist of all free cumulants beyond quadratic order in the fields,
\[
    G_{\underbracket[.5pt]{\scriptstyle f\dots f}_{n_f} \underbracket[.5pt]{\scriptstyle\mathcal{B}\dots\mathcal{B}}_{n_\B}}^{(0)} &\equiv\!
    \begin{tikzpicture}[baseline=(current bounding box.center)]
        \draw[thick, -{Latex[length=3mm/2,width=3mm/2]}] (0,0) -- (-1/2,-1/2);
        \draw[thick, -{Latex[length=3mm/2,width=3mm/2]}] (0,0) -- (-1/2,1/2);
        \draw[thick, -{Latex[reversed,length=3mm/2,width=3mm/2]}] (0,0) -- (1/2,1/2);
        \draw[thick, -{Latex[reversed,length=3mm/2,width=3mm/2]}] (0,0) -- (1/2,-1/2);
        \filldraw (0,0) circle (.05);
        \filldraw (-1/2,.3/2) circle (.05/3);
        \filldraw (-1/2,0) circle (.05/3);
        \filldraw (-1/2,-.3/2) circle (.05/3);
        \filldraw (1/2,.3/2) circle (.05/3);
        \filldraw (1/2,0) circle (.05/3);
        \filldraw (1/2,-.3/2) circle (.05/3);
        \node[left] at (-1/2,0) {$n_f$};
        \node[right] at (1/2,0) {$n_\B$};
    \end{tikzpicture}.
\]

In terms of these Feynman rules, a loop expansion scheme dubbed \emph{macroscopic perturbation theory} is established. As usual in QFT or other statistical field theories, $S_\V$ can be turned into an operator and the remaining Gaussian path integral can be solved,
\[\label{eq:marco_pt}
    \Z[J] &= \exp(S_\V\qty[\fdv{J}]) \int\DD\psi \exp(-\frac{1}{2}\psi^\transp\,\Delta^{-1}\,\psi + J^\transp\psi)\\
    &= \exp(S_\V\qty[\fdv{J}])\exp(\frac{1}{2} J^\transp\Delta J).
\]
Cumulants of the full theory are again generated by appropriate functional derivatives of the Schwinger functional $W[J] = \ln \Z[J]$.

\section{Application to cosmology}
The formalism presented in the previous section is valid for any classical $N$-particle system governed by a Hamiltonian of the form~\eqref{eq:hamiltonian}. There are two things that need to be specified in order to apply the theory to a specific system: an interaction potential and an initial phase-space density of the ensemble.

For cosmic structure formation, the interaction potential is obtained from relativistic perturbation theory around a spatially homogeneous and isotropic Friedmann-Lemaître-Robertson-Walker background. Sufficiently below the horizon and for cold dark matter, it corresponds to the generalized Newtonian gravitational potential. We will use the time coordinate $\eta = \ln D_+/D_+^\ini$, in which the potential is approximately time-independent (see Appendix~\ref{app:time_and_potential}). In terms of comoving coordinates, the potential energy of a point source then reads
\[\label{eq:point_source_potential}
    v(k) = -\frac{3}{2}\frac{1}{\bar{\rho}\,k^2},
\]
in Fourier space, with the mean particle number density~$\bar{\rho}$. We choose $z^\ini=1100$ and a Planck-like $\Lambda$CDM cosmology with $\Om=0.311$, $\Ol=1-\Om$, $h=0.677$, $\sigma_8=0.81$ and $n_s=0.967$~\cite{Planck2018}.

In the remaining part of this section, we will discuss different choices of sampling the initial phase-space distribution from a given density power spectrum. We obtain the latter from the Boltzmann solver CAMB~\cite{CAMB}.

\subsection{Initial conditions in earlier works}
\label{sec:old_ICs}
In the microscopic picture, with discrete particles, an initial phase-space density for cosmic structure formation has been derived in~\cite{Fabis2015,Bartelmann2016}, and we sketch the key steps in the following.
We assume the initial density field $\delta^\ini$ and the initial velocity field $\vec{v}^\ini$ to be Gaussian random fields with covariance
\[
    \mat{C}(\vec{q}_1,\vec{q}_2) = \begin{pmatrix}
    C_{\delta\delta}(\vec{q}_1,\vec{q}_2) & \vec{C}_{\delta \vec{v}}^\transp(\vec{q}_1,\vec{q}_2) \\
    \vec{C}_{\vec{v}\delta}(\vec{q}_1,\vec{q}_2) & \mat{C}_{\vec{v}\otimes\vec{v}}(\vec{q}_1,\vec{q}_2)
    \end{pmatrix}.
\]
The mean values of $\delta$ and $\vec{v}$ vanish by definition and isotropy, respectively, and due to homogeneity and isotropy, $\mat{C}$ only depends on the relative separation $\vec{r}=\vec{q}_1-\vec{q}_2$ and $\vec{C}_{\vec{v}\delta}(\vec{r}) = -\vec{C}_{\delta \vec{v}}(\vec{r})$.

We can then Poisson-sample each particle's initial position from the density field and use the velocity field to assign their momenta, i.\,e.~employing the conditional probability
\[\label{eq:P_cond}
    P(\vec{q}_j,\vec{p}_j \mid \delta^\ini, \vec{v}^\ini] = \frac{1}{V} \qty(1+\delta^\ini(\vec{q}_j)) \, \dirac\qty(\vec{p}_j-m\vec{v}^\ini(\vec{q}_j)).
\]
As described in Appendix~\ref{app:old_ICs}, the resulting initial phase-space density of the ensemble has the form
\begin{multline}
    \label{eq:N_particle_old}
    \varrho_N(\bm{x}^\ini) = \frac{V^{-N}}{\sqrt{\det 2\pi m^2\, \mathbf{C}_{\bm{v\otimes v}}}}\, \mathcal{C}(-\im\bm{\nabla}_{\bm{p}^\ini})\\\times\exp(-\frac{1}{2 m^2} {\bm{p}^\ini}^\transp \mathbf{C}_{\bm{v\otimes v}}^{-1}\,\bm{p}^\ini),
\end{multline}
with some polynomial $\mathcal{C}$.
In particular, for $N=1$ the phase space density reads
\[\label{eq:rho_1}
\varrho_1(p) = \frac{V^{-1}}{(2\pi \sigma_p^2)^{\frac{3}{2}}}\exp(-\frac{p^2}{2\sigma_p^2}),
\]
with the momentum variance $\sigma_p^2 = \frac{m^2}{3}\tr \mat{C}_{\vec{v} \otimes \vec{v}}(0)$, and for $N=2$ the phase-space density is
\begin{multline}
    \label{eq:rho_2}
    \varrho_2(\vec{r},\bm{p}) = \frac{m^6}{V^2}\int\frac{\dd\bm{l}_{\bm{v}}}{(2\pi)^6} \e^{\im \bm{l}_{\bm{v}}\cdot\frac{\bm{p}}{m}} \exp(-\frac{1}{2} \bm{l}_{\bm{v}}^\transp \mathbf{C}_{\bm{v\otimes v}}\,\bm{l}_{\bm{v}})\\
    \times\qty[ C_{\delta\delta}(\vec{r})
    + \qty(1-\im \vec{C}_{\delta\vec{v}}(\vec{r})\cdot\vec{l}_{v_1})
    \qty(1+\im\vec{C}_{\delta\vec{v}}(\vec{r})\cdot\vec{l}_{v_2})].
\end{multline}
If the initial time is chosen early enough, the single-stream approximation provides an accurate description of perfectly cold dark matter. The density contrast $\delta$ and the velocity field $\vec{v}$ are then described by the Euler-Poisson system of equations. As long as $\delta$ and $\abs{\vec{v}}$ are small, the latter can be linearized.
Note that the velocity~$\vec{v}$ with respect to our time coordinate~$\eta$ is related to the usual comoving velocity~$\vec{u}$ via
\[\label{eq:velocity_relation}
\vec{v} = \frac{\vec{u}}{H f_+},
\]
with the Hubble function $H$ and $f_+ = \dv*{\ln D_+}{\ln a}$ (see Appendix~\ref{app:time_and_potential}).

In the linear approximation, the growing mode of the density contrast evolves as $\delta(\vec{q},\eta) = \e^\eta\,\delta^\ini(\vec{q})$. Furthermore, any nonvanishing initial vorticity would decay and can therefore be ignored~\cite{Bernardeau2002}. In Fourier space, we thus have
\[\label{eq:v_ini}
    \vec{v}^\ini(\vec{k}) = \im \frac{\vec{k}}{k^2} \delta^\ini(\vec{k}),
\]
by the linearized continuity equation.
It is then sufficient to specify the initial density power spectrum to completely fix the joint covariance matrix $\mat{C}$.

\subsubsection{Cumulants and shot noise}
In accordance with~\eqref{eq:ev_observables}, the initial mean phase-space density or, equivalently, its one-point cumulant, reads
\[
    G_f^\ini(x) = \ev{f(x,\eta=0)} = \int\dd\bm{x}^\ini\,\varrho_N(\bm{x}^\ini) f(x,\eta=0).
\]
As per the definition~\eqref{eq:klimontovich} of $f$, it is just a sum of $N$ Dirac distributions. In each term, one of the $x_j^\ini$-integrals will resolve the Dirac distribution while the remaining $N-1$ integrals only act on~$\varrho_N$. Since the latter is obtained from Poisson-sampling a multivariate Gaussian [cf.~\eqref{eq:P_conditional}], this will simply marginalize to $\varrho_1$ given in~\eqref{eq:rho_1}. Hence,
\[
    G_f^\ini(x) = N \varrho_1(p).
\]
Trivially, we find
\[\label{eq:one_pt_density_vPi}
    G_\rho^\ini(x) = \bar{\rho}, \qq{and} G_{\vPi}^\ini(x) = 0,
\]
due to homogeneity and isotropy.
The initial two-point cumulant reads
\begin{multline}
    G_{ff}^\ini(x_1,x_2) = N(N-1)\,\varrho_2(\vec{q}_1-\vec{q}_2,\vec{p}_1,\vec{p}_2) \\- G_f^\ini(x_1)\,G_f^\ini(x_2)
    + N \varrho_1(p_1)\,\dirac(x_1-x_2).
\end{multline}
Since $\varrho_N \propto V^{-N}$, the first line has a term proportional to $\bar{\rho}^2$ and one proportional to $\bar{\rho}/V$. In the ``thermodynamic'' limit $N,V\to\infty$ with $\bar{\rho}=\text{const.}$, the latter can be neglected. The last term in the second line is proportional to $\bar{\rho}$. If the scales $r$ at which we evaluate the correlation function are much larger than the mean particle separation, which is the case for cosmic structure formation, we have $r^3\bar{\rho} \gg 1$. Therefore, we can ignore all but the dominant order in the mean density. The other terms are called \emph{shot noise}~\cite{Fabis2015,Bartelmann2016}. As expected, the two-point cumulant of the density is simply $G_{\rho\rho}^\ini = \bar{\rho}^2 C_{\delta\delta}$.

\subsubsection{Inconsistencies}
While the density cumulants trivially reproduce the expected expressions, inconsistencies appear for statistics involving higher momentum moments. In particular, the mean phase-space density $G_f^\ini$ contains the nonvanishing velocity dispersion $\sigma_p^2$, even though we started from perfectly cold dark matter. Moreover, the initial two-point function of the momentum density reads
\[\label{eq:initial_momenta_old}
    \mat{G}_{\vPi\otimes\vPi}^\ini = \bar{\rho}_m^2\qty[(1+C_{\delta\delta})\mat{C}_{\vec{v}\otimes\vec{v}} + \vec{C}_{\delta\vec{v}} \otimes \vec{C}_{\vec{v}\delta}].
\]
The quadratic terms follow from the nonlinear relation $\vec{\Pi} = \bar{\rho}_\mathrm{m}(1+\delta)\vec{v}$ between the momentum density and the density and velocity fields. In fact, already~the conditional probability~\eqref{eq:P_cond} is nonlinear in the velocity field and consequently (\ref{eq:N_particle_old}--\ref{eq:rho_2}) contain arbitrarily high orders in the latter.
However, in order to set the initial conditions, we have linearized the Euler-Poisson system in $\delta$ and $\vec{v}$. Therefore, only the leading order terms should have been kept in order to be consistent.
Furthermore, if all terms are kept, Gaussian initial conditions in $\delta$ and $\vec{v}$ lead to non-Gaussianities in~$\vPi$. Indeed, all higher $n$-point cumulants for the phase-space density are nonvanishing,
\[
    G_{f\dots f}^\ini(x_1,\dots,x_n) ={}& \frac{N!}{(N-n)!}\,\varrho_n(x_1,\dots,x_n)\\
    &({}-\text{disconn.~parts} + \text{shot noise}),
\]
with $N!/(N-n)! \approx N^n$, even if shot noise is neglected. Had one only kept the leading order terms in the initial phase-space density, this would not be the case.

In summary, the inconsistent linearization introduces spurious effects like a velocity dispersion, higher-order contributions to initial cumulants and non-Gaussianities.
Nevertheless, these initial conditions have been used for both the microscopic and macroscopic approach to KFT~\cite{Fabis2015,Bartelmann2016,Fabis2018,Lilow2018,Lilow2019,Pixius2022,Heisenberg2022,Konrad2022,Daus2025}.
Even though the spurious effects are initially small, they make the evaluation even of freely evolved phase-space density cumulants difficult\footnote{The main difficulty are oscillatory integrals of the form~\eqref{eq:rho_2}. For two-point functions, they can be Fourier-transformed, integrated with standard techniques and afterward transformed back~\cite{Daus2025}, e.\,g.\ using FFTLog~\cite{Talman1978, Hamilton2000, Gebhardt2018}. For higher-point functions, this is no longer possible.}, prevent an analytical solution of the macroscopic propagator~\cite{Lilow2018,Lilow2019}, and impact late-time observables on small scales, as we will discuss in Sec.~\ref{sec:tree_level}.

In fact, sampling from the configuration space density and thereafter enforcing a velocity associated to the position does not treat positions and momenta on equal footing. This motivates us to choose different initial conditions directly in the phase-space density, which employ the linearization consistently.

\subsection{Perfectly cold initial conditions}
\label{sec:new_ICs}
In the single-stream approximation, which we assume to hold \emph{only at the initial time}, the phase-space density is of the form
\[
    f^\ini(\vec{q},\vec{p}) = \bar{\rho}\,\qty(1+\delta^\ini(\vec{q}))\,\dirac\qty(\vec{p}-m\vec{v}^\ini(\vec{q})).
\]
In Fourier space, linearizing in $\delta^\ini$ and $\vec{v}^\ini$ yields
\[
    f^\ini(\vec{k},\vec{l}) &\approx (2\pi)^3\dirac(\vec{k})\,\bar{\rho}+\bar{\rho}\,\delta^\ini(\vec{k})-\im\,\vec{l}\cdot \vec{v}^\ini(\vec{k})\,\bar{\rho} \\
    &= (2\pi)^3\dirac(\vec{k})\,\bar{\rho}+\qty(1+\frac{\vec{k}\cdot\vec{l}}{k^2})\,\bar{\rho}\,\delta^\ini(\vec{k}),
\]
where $\vec{k}$ and $\vec{l}$ are the Fourier conjugates to $\vec{q}$ and $\vec{p}/m$, respectively, and we set the growing-mode initial conditions~\eqref{eq:v_ini}. Given this linear relationship between $f^\ini$ and the Gaussian random field $\delta^\ini$, the former is also Gaussian with mean and variance
\[
    \mu_f(\vec{k}_1,\vec{l}_1) ={}& (2\pi)^3\dirac(\vec{k}_1)\,\bar{\rho}\\
    C_{ff}(\vec{k}_1,\vec{l}_1,\vec{k}_2,\vec{l}_2) ={}& (2\pi)^3 \dirac(\vec{k}_1+\vec{k}_2)\,\bar{\rho}^2\,P_\delta^\ini(k_1)\\
    &\times \qty(1+\frac{\vec{k}_1\cdot\vec{l}_1}{k_1^2})\qty(1-\frac{\vec{k}_1\cdot\vec{l}_2}{k_1^2}).
\]
An initial phase-space density for the particle ensemble can be obtained by Poisson-sampling in phase space (see Appendix~\ref{app:new_ICs}).
As expected, the resulting cumulants are
\[
    G_f^\ini = \mu_f \qq{and} G_{ff}^\ini = C_{ff}\ (+\ \text{shot noise}),
\]
all higher cumulants are pure shot noise and vanish in the thermodynamic limit.

\subsection{Tree-level spectra}
\label{sec:tree_level}
\begin{figure*}
    \includegraphics[width=.5\linewidth]{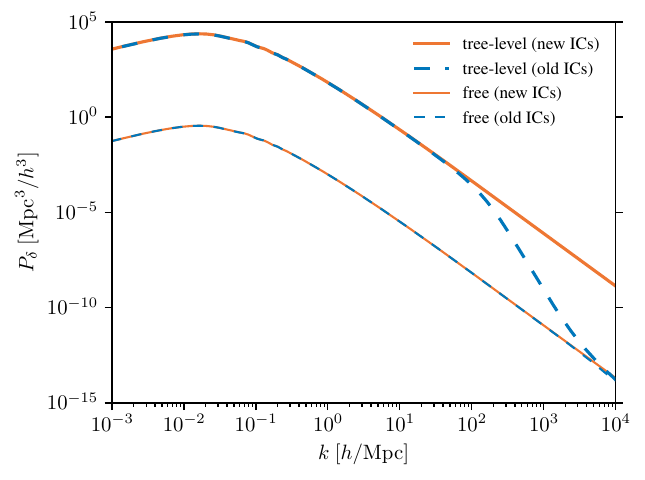}\hfill
    \includegraphics[width=.5\linewidth]{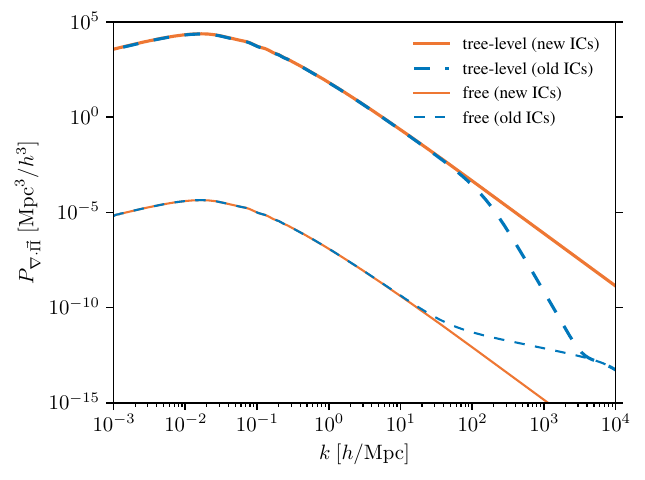}
    \caption{\label{fig:tree_level}
        Tree-level power spectra for the density contrast (left panel) and the divergence of the momentum density (right panel). Solid orange lines represent the results from the newly proposed initial conditions~\ref{sec:new_ICs}, while dashed blue lines correspond to the initial conditions~\ref{sec:old_ICs} used in previous works. For reference, the corresponding freely evolved power spectra are shown as thin lines of the same style. For dimensional consistency, note the relation~\eqref{eq:velocity_relation}.
    }
\end{figure*}

As detailed in Appendix~\ref{app:macro_details}, the retarded macroscopic propagator~\eqref{eq:Neumann_series} can be obtained by solving the Volterra integral equation
\[
    \widetilde{\Delta}_{f\B}(\vec{k},\vec{l},\eta_1,\eta_2) ={}& G_{f\B}^\free(\vec{k},\vec{l},\eta_1,\eta_2) \\
    &+ \int\dd \eta\,G_{f\B}^\free(\vec{k},\vec{l},\eta_1,\eta)\,\widetilde{\Delta}_{f\B}(\vec{k},0,\eta,\eta_2).
\]
For the initial conditions~\ref{sec:old_ICs}, this equation can only be solved numerically~\cite{Lilow2019}. However, for the initial conditions~\ref{sec:new_ICs}, an analytical solution~\eqref{eq:Laplace_solution} can be obtained using Laplace transformations. It reads
\begin{multline}
    \widetilde{\Delta}_{f\B}(\vec{k},\vec{l},\eta_1,\eta_2) = \frac{3}{5} \Bigg[\qty(1+\frac{\vec{k}\cdot\vec{l}}{k^2})\,\e^{\eta_1-\eta_2} \\
    - \qty(1-\frac{3}{2}\frac{\vec{k}\cdot\vec{l}}{k^2})\,\e^{-\frac{3}{2}(\eta_1-\eta_2)}\Bigg]\Theta(\eta_1-\eta_2).
\end{multline}
By~\eqref{eq:def_resummed_propagator}, the tree-level propagator is
\begin{multline}
    \label{eq:tree_level_result}
    \Delta_{ff}(\vec{k}_1,\vec{l}_1,\eta_1,\vec{k}_2,\vec{l}_2,\eta_2) = (2\pi)^3\dirac(\vec{k}_1+\vec{k}_2)\,\bar{\rho}^2\,\e^{\eta_1+\eta_2}
    \\
    \times\qty(1+\frac{\vec{k}_1\cdot\vec{l}_1}{k_1^2}) \qty(1-\frac{\vec{k}_1\cdot\vec{l}_2}{k_1^2})\,P_\delta^\ini(k_1).
\end{multline}
The power spectra of the density contrast and the momentum density (which is equal to the velocity field in the linear approximation) are exactly those of linear SPT with growing-mode initial conditions. In Fig.~\ref{fig:tree_level}, we show the tree-level density and momentum-divergence power spectra for both sets of initial conditions.
On large and intermediate scales, the tree-level spectra agree with each other. On very small scales, $k\gtrsim 50\,h\,\unit{Mpc^{-1}}$, however, the spectra are damped for the initial conditions from~\ref{sec:old_ICs}, until they reach the amplitude of the freely evolved spectra. This is consistent with the findings of~\cite{Lilow2019,Kozlikin2021}. 
However, since the underlying dynamics are identical in both cases, this effect can be completely attributed to the method of sampling the initial conditions. In fact, the damping is a consequence of the exponential in~\eqref{eq:N_particle_old}, which is in turn rooted in the inconsistent linearization in the initial fields and has the effect of an artificial velocity dispersion. Spurious mode-coupling terms caused by the aforementioned inconsistency can already be seen in the freely evolved momentum power spectra. This not only includes the small quadratic corrections discussed for~\eqref{eq:initial_momenta_old}, but also contributions which grow with time.
The aforementioned effects are distinct from decaying contributions generated by small deviations in the initial conditions that are also known as transients~\cite{Scoccimarro1998a,Crocce2006a}, as both sampling methods single out the growing modes. Whereas the free evolution is treated in an exact manner in RKFT, gravitational interactions are only included perturbatively. For that reason, even seemingly subdominant inaccuracies in the initial conditions can lead to nonvanishing effects at any finite perturbative order.

Having demonstrated that the tree-level results of RKFT agree with SPT when consistent initial conditions are chosen, we direct our attention to the first loop order in the following, restricting to the newly proposed sampling method.

\subsection{One-loop spectra}
\label{sec:one_loop}
In Appendix~\ref{app:self_energies}, we list the nine Feynman diagrams contributing to the one-loop spectra. Since we are interested in late-time statistics, we limit to the contributions that grow fastest with time, as usual in SPT~\cite{Bernardeau2002}. For the equal-time power spectra, these stem from the three diagrams
\[
    \frac{1}{2}\,\begin{tikzpicture}[baseline=-.75mm]
    \draw[thick, {Latex[length=3mm/2,width=3mm/2]}-] (.75/4,0) -- (.5,0);
    \draw[thick] (0,0) -- (.5,0);
    \draw[thick, -{Latex[length=3mm/2,width=3mm/2]}] (1.5,0) -- (1.25/4+1.5,0);
    \draw[thick] (1.5,0) -- (2,0);
    \draw[thick] (1,0) circle (.5);
    \filldraw (.5,0) circle (.05);
    \filldraw (1.5,0) circle (.05);
    \filldraw (1,.5) circle (.05);
    \filldraw (1,-.5) circle (.05);
    \draw[thick, -{Latex[length=3mm/2,width=3mm/2]}] (1-.5/1.414+.001,.5/1.414+.0008)--(1-.5/1.414,.5/1.414);
    \draw[thick, -{Latex[length=3mm/2,width=3mm/2]}] (1+.5/1.414-.001,.5/1.414+.0008)--(1+.5/1.414,.5/1.414);
    \draw[thick, -{Latex[length=3mm/2,width=3mm/2]}] (1-.5/1.414+.001,-.5/1.414-.0008)--(1-.5/1.414,-.5/1.414);
    \draw[thick, -{Latex[length=3mm/2,width=3mm/2]}] (1+.5/1.414-.001,-.5/1.414-.0008)--(1+.5/1.414,-.5/1.414);
    \end{tikzpicture}
    + 2\,\begin{tikzpicture}[baseline=-.75mm]
    \draw[thick, {Latex[length=3mm/2,width=3mm/2]}-] (.75/4,0) -- (.5,0);
    \draw[thick] (0,0) -- (.5,0);
    \draw[thick] (1,0) circle (.5);
    \filldraw (.5,0) circle (.05);
    \filldraw (1.5,0) circle (.05);
    \filldraw (1,.5) circle (.05);
    \draw[thick, -{Latex[length=3mm/2,width=3mm/2]}] (1-.5/1.414+.001,.5/1.414+.0008)--(1-.5/1.414,.5/1.414);
    \draw[thick, -{Latex[length=3mm/2,width=3mm/2]}] (1+.5/1.414-.001,.5/1.414+.0008)--(1+.5/1.414,.5/1.414);
    \draw[thick, -{Latex[length=3mm/2,width=3mm/2]}] (1,-.5)--(1-.001-.1,-.5);
    \draw[thick, {Latex[length=3mm/2,width=3mm/2]}-{Latex[length=3mm/2,width=3mm/2]}] (1.625,0) -- (1.5+1.75/2,0);
    \draw[thick] (1.5,0) -- (2.5,0);
    \filldraw (2,0) circle (.05);
    \end{tikzpicture}
    + \begin{tikzpicture}
    \draw[thick, {Latex[length=3mm/2,width=3mm/2]}-] (.75/2,0) -- (1,0);
    \draw[thick] (0,0) -- (1,0);
    \filldraw (1,0) circle (.05);
    \draw[thick, {Latex[length=3mm/2,width=3mm/2]}-{Latex[length=3mm/2,width=3mm/2]}] (1.25+.25/2,0) -- (1.25+1.75/2,0);
    \draw[thick] (1,0) -- (2.25,0);
    \filldraw (1.75,0) circle (.05);
    \draw[thick] (1,.5) circle (.5);
    \filldraw (1,1) circle (.05);
    \draw[thick, {Latex[length=3mm/2,width=3mm/2]}-] (.5,.449)--(.5,.451);
    \draw[thick, {Latex[length=3mm/2,width=3mm/2]}-] (1.5,.449)--(1.5,.451);
    \end{tikzpicture}.
\]

\begin{figure}
    \includegraphics[width=\linewidth]{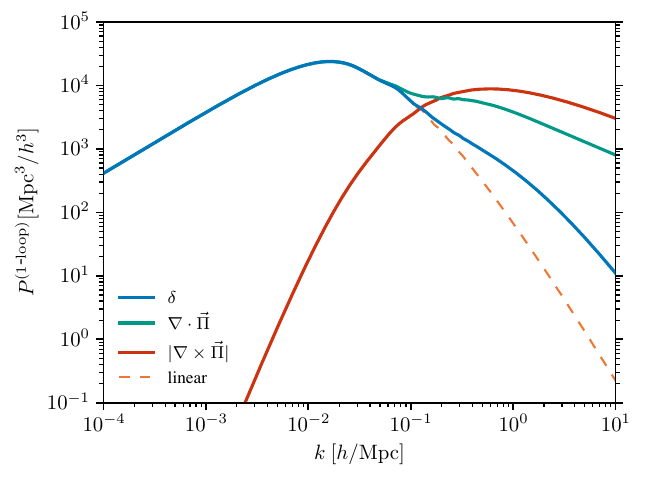}
    \caption{\label{fig:one_loop}
        One-loop power spectra of the density contrast (blue), divergence of the momentum density (green), and the absolute value of its curl (red). The linear power spectrum of $\delta$ (or, equivalently, $\nabla\cdot\vPi$) is shown as a dashed orange line. For dimensional consistency, note the relation~\eqref{eq:velocity_relation}.
    }
\end{figure}

In Fig.~\ref{fig:one_loop}, we show the one-loop power spectra of the density contrast, the divergence of the momentum density and its curl.
Note that we compute spectra of the momentum density $\vPi=\rho\vec{v}$ and not velocity spectra\footnote{In contrast to the divergence and curl of the velocity field, $\nabla\cdot\vec{\Pi}$ and $\nabla\times\vec{\Pi}$ are not invariants under Galilean transformations. Some care has to be taken when connecting them to observables.}, since the former is a more natural object in KFT and useful for observational purposes~\cite{Park2000,Park2006,Park2016}.
In particular, the nonvanishing curl power spectrum is \emph{not} proof of vorticity $\vec{\omega}=\nabla\times\vec{v}$, since
\[
    \nabla\times(\rho\vec{v}) = \nabla\rho\times\vec{v} + \rho\vec{\omega}.
\]

\noindent The results of all spectra agree exactly with one-loop~SPT,
\[
    \Delta P_\delta^\text{(1-loop)}(k,\eta) ={}& P_{13}(k,\eta) + P_{22}(k,\eta)\\
    \Delta P_{\nabla\cdot\vPi}^\text{(1-loop)}(k,\eta) ={}& 3\,P_{13}(k,\eta) + 4\,P_{22}(k,\eta)\\    
    P_{\abs{\nabla\times\vPi}}^\text{(1-loop)}(k,\eta) ={}& 
    \e^{4\eta}\int\frac{\dd[3]{k'}}{(2\pi)^3} P_\delta^\ini(k')\,P_\delta^\ini(\abs*{\vec{k}-\vec{k}'}) \\ &\times\abs{\vec{k}\times\vec{k}'}^2\qty(\frac{1}{k'^2}-\frac{1}{|\vec{k}-\vec{k}'|^2})^2,
\]
where $P_{13}$ and $P_{22}$ are the usual SPT one-loop corrections~\cite{Bernardeau2002}. The prefactors in the momentum divergence power spectrum follow directly from the time derivative in the continuity equation. From the equivalence of the curl power spectrum, we infer that there is no vorticity contribution.
In~\cite{Lilow2018}, the one-loop density-fluctuation power spectrum was computed for the initial conditions~\ref{sec:old_ICs} for the first time. However, due to the computational difficulties in evaluating the corresponding free cumulants, they were expanded in powers of the initial power spectrum. The calculation  without this expansion was done in~\cite{Daus2025}. Again, their result differs from ours by a damping on very small scales, which can be attributed to the choice of initial conditions as discussed in the previous section.

The stress tensor $\mathrm{T}$ defined in~\eqref{eq:momentum_moments} can be decomposed as
\[
    \mathrm{T} = (1+\delta)\bar{\rho}\,(\vec{v}\otimes\vec{v} + \sigma),
\]
with the $\rank 2$ velocity dispersion tensor $\sigma$. The latter is set to zero in SPT in order to close the moment hierarchy of the Vlasov equation. This is the \emph{single-stream approximation}. In KFT, it is \emph{a priori} possible to have a nonvanishing velocity dispersion tensor, since no truncation of the Vlasov hierarchy was assumed. In fact, it should build up dynamically in order to be compatible with the generation of vorticity observed in simulations~\cite{Pueblas2009,Hahn2015}.
The lowest order contribution to $\mathrm{T}$ in SPT is the term quadratic in $\vec{v}$, which is why a first comparison to KFT can be done at one-loop order.
Again, we restrict to the contributions that grow fastest in time. The one-point function agrees,
\[
    \mathrm{G}_\mathrm{T}^\text{(1-loop)}(\vec{k},\eta) = (2\pi)^3\dirac(\vec{k})\,\bar{\rho}\,\e^{2\eta}\,\sigma_p^2\identity_3.
\]
It is no surprise that it is diagonal, i.\,e.~that it cannot contain anisotropic stress contributions, since it is an averaged quantity. In fact, $\mathrm{G}_\mathrm{T}$ agrees for SPT and KFT even at two-loop level, involving contributions which are of fourth order in the fields. The more interesting quantities are therefore higher-point cumulants.

\begin{widetext}
    The late-time contributions to the $n$-point function comes from diagrams of the form
    \[
        \mathrm{G}_{\mathrm{T}\dots\mathrm{T}}^\text{(1-loop)} = \frac{n!}{(-2)^n} (\nabla_{l_1}\odot\nabla_{l_1})\odot\dots\odot(\nabla_{l_n}\odot\nabla_{l_n})
        \begin{tikzpicture}[baseline=-.75mm]
            \draw[thick] (0,0) -- (135:1) node[above left=-1mm] {$n$};
            \draw[thick] (0,0) -- (45:1) node[above right=-1mm] {$1$};
            \draw[thick] (0,0) -- (-45:1) node[below right=-1mm] {$2$};
            \draw[thick] (0,0) -- (-135:1) node[below left=-1mm] {$3$};
            \filldraw[thick, fill=white] ([shift={(150:.5)}]0,0) arc (150:-150:.5);
            \draw[thick, dotted] ([shift={(-150:.5)}]0,0) arc (-150:-210:.5);
            \filldraw (-135:0.5) circle (.04);
            \filldraw (-45:0.5) circle (.04);
            \filldraw (45:0.5) circle (.04);
            \filldraw (135:0.5) circle (.04);
            \filldraw (-90:0.5) circle (.04);
            \filldraw (0:0.5) circle (.04);
            \filldraw (90:0.5) circle (.04);
            \draw[thick, -{Latex[length=1mm,width=3mm/2]}] (-135:.8) -- (-135:.85);
            \draw[thick, -{Latex[length=1mm,width=3mm/2]}] (-45:.8) -- (-45:.85);
            \draw[thick, -{Latex[length=1mm,width=3mm/2]}] (45:.8) -- (45:.85);
            \draw[thick, -{Latex[length=1mm,width=3mm/2]}] (135:.8) -- (135:.85);
            \draw[thick, -{Latex[length=1mm,width=3mm/2]}] (-90-25:.5) -- (-90-30:.5);
            \draw[thick, -{Latex[length=1mm,width=3mm/2]}] (-90+25:.5) -- (-90+30:.5);
            \draw[thick, -{Latex[length=1mm,width=3mm/2]}] (-25:.5) -- (-30:.5);
            \draw[thick, -{Latex[length=1mm,width=3mm/2]}] (+25:.5) -- (+30:.5);
            \draw[thick, -{Latex[length=1mm,width=3mm/2]}] (90-25:.5) -- (90-30:.5);
            \draw[thick, -{Latex[length=1mm,width=3mm/2]}] (90+25:.5) -- (90+30:.5);
        \end{tikzpicture}
        \Bigg|_{\vec{l}_1=\dots=\vec{l}_n=0},
    \]
    where $\odot$ denotes the symmetric tensor product. Since the tree-level propagator~\eqref{eq:tree_level_result}, $\Delta_{ff}=
    \begin{tikzpicture}
        \draw[thick, {Latex[length=3mm/2,width=3mm/2]}-{Latex[length=3mm/2,width=3mm/2]}] (.25/2,0) -- (1.75/2,0);
        \draw[thick] (0,0) -- (1,0);
        \filldraw (.5,0) circle (.05);    
    \end{tikzpicture}$, factorizes, this diagram can be decomposed into factors of the form
    \[
        -\frac{1}{2}(\nabla_{l}\odot\nabla_{l})\,
        \begin{tikzpicture}[baseline=-.75mm]
            \draw[thick] (0,0) -- (180:.5);
            \draw[thick] (0,0) -- (45:.49);
            \draw[thick] (0,0) -- (-45:.49);
            \filldraw (0,0) circle (.05);
            \draw[thick, fill=white] ([shift={(45:.5)}]-45:.05) arc (-45:-225:.05);
            \draw[thick, fill=white] ([shift={(-45:.5)}]45:.05) arc (45:225:.05);
            \draw[thick, -{Latex[length=1mm,width=3mm/2]}] (180:.3) -- (180:.35);
            \draw[thick, -{Latex[length=1mm,width=3mm/2]}] (45:.25) -- (45:.2);
            \draw[thick, -{Latex[length=1mm,width=3mm/2]}] (-45:.25) -- (-45:.2);
        \end{tikzpicture}
        \Bigg|_{\vec{l}=0}
        = -\bar{\rho}\,\e^{2\eta} \int_{\vec{k}_2,\vec{k}_3} (2\pi)^3\dirac(\vec{k}-\vec{k}_2-\vec{k}_3) \qty(\frac{\vec{k}_2}{k_2^2}\odot\frac{\vec{k}_3}{k_3^2}),
    \]
    yielding
    \[
        \mathrm{G}_{\mathrm{T}\dots\mathrm{T}}^\text{(1-loop)}(\vec{k}_1,\eta_1,\dots,\vec{k}_n,\eta_n) = (2\pi)^3\dirac\qty(\sum_{i=1}^{n}\vec{k}_i)\,\bar{\rho}^n n!\,\exp(2\sum_{i=1}^{n}\eta_i) \int_{\vec{k}} \bigodot_{i=1}^n\left[\frac{\qty(\vec{k}-\sum_{j=1}^{i-1}\vec{k}_j)^{\odot2}}{\abs{\vec{k}-\sum_{j=1}^{i-1}\vec{k}_j}^4} P_\delta^\ini\qty(\abs{\vec{k}-\sum_{j=1}^{i-1}\vec{k}_j})\right].
    \]
    This is identical to the SPT result,
    \[
        \ev{(\bar{\rho}\vec{u}^{(1)}\odot\vec{u}^{(1)})(\vec{k}_1,\eta_1)\odot\dots\odot(\bar{\rho}\vec{u}^{(1)}\odot\vec{u}^{(1)})(\vec{k}_n,\eta_n)}_\mathrm{c} = \mathrm{G}_{\mathrm{T}\dots\mathrm{T}}^\text{(1-loop)}(\vec{k}_1,\eta_1,\dots,\vec{k}_n,\eta_n).
    \]
    Having shown that all $n$-point cumulants of the stress tensor agree at one-loop, we conclude that no velocity dispersion tensor $\sigma$ builds up at that order.
\end{widetext}

\section{Summary, conclusions and outlook}
\label{sec:conclusion}
In this paper, we investigate cosmic structure formation in the framework of a path-integral description of a classical $N$-particle ensemble. In particular, we use perturbation theory in a macroscopic reformulation of the theory, in which gravitational interactions are partly resummed (without introducing free effective parameters).
In Sec.~\ref{sec:old_ICs}, we review the initial conditions that were used in prior work, which inconsistently employ the linearized continuity equation in order to relate the statistics of the velocity field to those of the density contrast. As a result, the initial $N$-particle phase-space density gets spurious higher-order contributions and becomes non-Gaussian. We propose an alternative way of setting initial conditions in Sec.~\ref{sec:new_ICs}, where the linearization is performed consistently. As a consequence, the initial phase-space density becomes Gaussian.

In Sec.~\ref{sec:tree_level}, we compare the tree-level power spectra of the density contrast and the divergence of the momentum density for both sets of initial conditions. For the newly proposed initial conditions, the tree-level propagator can be computed analytically and exactly reproduces the spectra of linear SPT. On the other hand, for the initial conditions of Sec.~\ref{sec:old_ICs}, the tree-level propagator needs to be evaluated numerically, and the resulting spectra are damped on very small scales, $k\gtrsim 50\,h\,\unit{Mpc^{-1}}$. The direct comparison shows that this effect is purely due to the differences in sampling the initial conditions. In particular, it does not allow the conclusion that the KFT perturbation theory captures effects that SPT does not. It is known from $N$-body simulations that small inaccuracies in the initial conditions can lead to transient effects. In our case, however, the nonperturbative treatment of the free evolution leads to non-negligible effects surviving at any finite perturbative order in the interactions.

In Sec.~\ref{sec:one_loop}, we investigate the one-loop corrections, focusing on the new initial conditions. Not only do the power spectra of the density contrast and the divergence and curl of the momentum density agree with the corresponding SPT spectra, but also all $n$-point cumulants of the stress tensor. At this order, therefore, neither a velocity dispersion tensor nor vorticity builds up. The two-loop expectation value of the stress tensor also matches its SPT counterpart. Since the one-loop stress tensor couples into the two-loop spectra, the vanishing one-loop velocity dispersion tensor suggests that the two-loop spectra will also match those of SPT.
Our results are consistent with earlier perturbative approaches to cosmic structure formation in the phase space based on the Vlasov (or collisionless Boltzmann) equation~\cite{Valageas2001,Nascimento2025}. There, equivalence with SPT has been found up to third order in the phase-space density. While we start from an $N$-particle ensemble, we neglect shot noise and therefore collision effects. In this limit, the dynamics of KFT and the Vlasov equation should indeed be equivalent.
In principle, the Vlasov dynamics allows for stream crossing even for perfectly cold initial conditions. For the cumulant hierarchy of the Vlasov equation, for which SPT poses a self-consistent truncation, however, this is no longer the case. Strictly speaking, the hierarchy becomes invalid at stream crossing, where the density contrast formally diverges~\cite{Pueblas2009}. Since perturbation theory with hierarchical initial conditions reproduces the Vlasov momentum cumulant hierarchy, the perturbative equivalence between the Vlasov-Poisson system (and therefore KFT) and SPT is a direct consequence of the self-consistency of the single-stream approximation. Although perfectly cold initial conditions had been assumed in previous KFT-related works, the inconsistent sampling introduced a spurious velocity dispersion, breaking the aforementioned fixed point and leading to differences compared to SPT. The present work thus clarifies the relation between KFT and conventional analytical approaches.

An obvious next step would be the inclusion of a \emph{physically motivated} initial velocity dispersion.
Even though the spectra will differ, however, we expect similar secularity problems as in SPT, i.\,e.~higher-order perturbative corrections growing faster in time, indicating a breakdown of perturbation theory~\cite{Carlson2009,Blas2014}. Therefore, it seems necessary to use nonperturbative techniques. Full phase-space approaches based on the Vlasov equation or (R)KFT provide a basis for exploring these without imposing an a priori truncation of the dynamics.

\begin{acknowledgments}
    We would like to thank Jendrik Marijan for proofreading and commenting on this manuscript.
    
    This work was supported by the Deutsche Forschungsgemeinschaft (DFG, German Research Foundation) under Germany's Excellence Strategy EXC 2181/1---390900948 (the Heidelberg STRUCTURES Excellence Cluster). H.H.~is supported by the Deutsche Forschungsgemeinschaft (DFG, German Research Foundation)---528166846.
    M.S.~is supported by Stiftung der Deutschen Wirtschaft (sdw).
\end{acknowledgments}

\section*{Data availability}
The data that support the findings of this article are openly available~\cite{Code}.

\appendix

\section{Derivation of the macroscopic action}
\label{app:rkft_derivation}
In order to bring the interacting part of the microscopic action~\eqref{eq:S_I_doublet} into the form of a kinetic term for the macroscopic fields, we can introduce a field doublet $\phi = (\B, f)^\transp$ and write it as quadratic term,
\[
\im S_\I[\phi] &= \frac{1}{2}\int\dd X_1\,\dd X_2 \, \phi(X_1)^\transp\,\sigma(X_1,X_2)\,\phi(X_2)\\
&\equiv \frac{1}{2} \phi^\transp\, \sigma\, \phi,
\]
with the matrix
\[
\sigma(X_1,X_2) \coloneq \begin{pmatrix}
    0 & \identity \\
    \identity & 0
\end{pmatrix}(X_1,X_2),
\]
where $\identity(X_1,X_2)\coloneq\dirac(X_1-X_2)$.
Having expressed the action $S_\I$ in terms of macroscopic fields, there are two ways of introducing path integrals over the latter into the partition function:
\begin{enumerate}
    \item Multiplying the integrand by unity \cite{Lilow2019},
    \[
    1 &= \int\DD\psi_f\, \dirac[\psi_f-f]\\
    &= \int\DD\psi_f\,\DD\psi_\B\, \e^{\im \psi_\B\cdot(\psi_f-f)},
    \]
    where the Dirac distribution ensures that $\psi_f$ equals the Klimontovich phase-space density $f$. In particular, $\psi_f$ will obey the same statistics as $f$.
    \item Performing a Hubbard-Stratonovich transformation \cite{Daus2024},
    \[
    \e^{\frac{1}{2} \phi^\transp\,\sigma\,\phi} \propto \int\DD\psi \exp(-\frac{1}{2}\psi^\transp\,\sigma^{-1}\,\psi + \psi^\transp\phi).
    \]
    Essentially, this is a Gaussian integral with source term over a newly introduced doublet field $\psi = (\psi_f, \psi_\B)^\transp$. The Hubbard-Stratonovich transformation is a widely used technique in many-body theory and particle physics. It allows the reformulation of a theory of particles interacting through two-body potentials in terms of a collective field that independent particles couple to.
\end{enumerate}
These two options are exactly equivalent\footnote{The original papers only differ in their convention as to whether to absorb $\im$ into $\B$ or not.}. The theory is now formulated in terms of the macroscopic fields of interest. In particular, it is a field theory in the doublet $\psi$, which contains $f$, of which $\rho$, $\vPi$ etc.~are momentum moments.
The full partition function takes the form
\[\label{eq:Z_mac_1}
\Z = \int\DD\psi \exp(-\frac{1}{2}\psi^\transp\,\sigma^{-1}\,\psi + \ln \Z^\free[\psi]).
\]
$\Z^\free[\psi]$ denotes the free generating functional with the macroscopic doublet $\psi$ acting as a source, i.\,e.\ $\Z^\free[\psi]$ is obtained from $\Z^\free$ via $\im S_0 \to \im S_0 + \psi\cdot\phi$.
The Schwinger functional $W^\free[\psi] = \ln\Z^\free[\psi]$ appearing in \eqref{eq:Z_mac_1} thus generates cumulants of the free theory, e.\,g.
\[\label{eq:GrB_def}
G_{f\B}^\free(X_1,X_2)
= \frac{\delta^2 W^\free[\psi]}{\delta\psi_\B(X_1)\,\delta\psi_f(X_2)} \Bigg\rvert_{\psi=0}.
\]
Indeed, it contains \emph{all possible} free $f$- and $\B-$field cumulants and lends itself to a cumulant expansion\footnote{Note that pure $\B$-field cumulants vanish \cite{Fabis2018}.}
\begin{multline}
    W^\free[\psi] = G_{f}^\free\cdot\psi_\B + \psi_\B\cdot G_{f\B}^\free \cdot\psi_f \\
    + \frac{1}{2} \psi_\B\cdot G_{f f}^\free \cdot\psi_\B + \mathcal{O}(\psi^3).
\end{multline}
In Appendix~\ref{app:free_cumulants}, we show how to explicitly compute them.
Collecting all quadratic terms, i.\,e.~all free two-point cumulants, in an inverse propagator $\Delta^{-1}$ and introducing a source doublet $J$, the generating functional of the full theory reads
\[
    \Z[J] = \int\DD\psi \exp(-\frac{1}{2}\psi^\transp\,\Delta^{-1}\,\psi + S_\V[\psi] + J\cdot\psi),
\]
where the vertex part of the action, $S_\V$, contains all remaining free cumulants of $f$ and $\B$.

\section{Free cumulants}
\label{app:free_cumulants}
Given an initial $n$-point phase-space cumulant $G_{f\dots f}^\ini$, the corresponding freely evolved cumulant is simply obtained by means of~\eqref{eq:Z_micro}, where the classical trajectory is replaced by the free motion, i.\,e.
\begin{multline}
    G_{f\dots f}^\free(\vec{q}_1,\vec{p}_1,t_1,\dots,\vec{q}_n,\vec{p}_n,t_n) \\
    = G_{f\dots f}^\ini\qty(\vec{q}_1-t_1\frac{\vec{p}_1}{m},\vec{p}_1,\dots,\vec{q}_n-t_n\frac{\vec{p}_n}{m},\vec{p}_n).
\end{multline}
It is often easier to work in Fourier space, especially in homogeneous systems. In that case, the above translates to\footnote{In the case of the time coordinate we use for the cosmological application (see \ref{app:time_and_potential}), replace $\vec{l}_i + t_i\frac{\vec{k}_i}{m} \mapsto \gpp(\eta_i)\,\vec{l}_i+\gqp(\eta_i)\frac{\vec{k}_i}{m}$.}
\begin{multline}
    G_{f\dots f}^\free(\vec{k}_1,\vec{l}_1,t_1,\dots,\vec{k}_n,\vec{l}_n,t_n) \\
    = G_{f\dots f}^\ini\qty(\vec{k}_1,\vec{l}_1+t_1\frac{\vec{k}_1}{m},\dots,\vec{k}_n,\vec{l}_n+t_n\frac{\vec{k}_n}{m}).
\end{multline}
Using the definition of the response field~\eqref{eq:response_field}, mixed $f$- and $\B$-field cumulants can be obtained from the above~\cite{Fabis2018}.
\begin{widetext}
    \noindent
    In particular, we get
    \[
    G_{f\B\dots\B}^\free(\vec{k}_1,\vec{l}_1,t_1,\dots,\vec{k}_n,\vec{l}_n,t_n) = (2\pi)^3 \dirac\qty(\sum_{i=1}^{n} \vec{k}_i) \prod_{i=2}^{n} \qty[(2\pi)^3\dirac(\vec{l}_i)\,v(\vec{k}_i,t_i)\,\vec{k}_i \cdot \vec{L}_n(t_i)] \mathcal{G}_{f}^\ini(\vec{L}_n(0)),
    \]
    in a homogeneous system, where $G_f^\ini(\vec{k},\vec{l}) = (2\pi)^3\dirac(\vec{k})\,\mathcal{G}_{f}^\ini(\vec{l})$. We defined
    \[
    \vec{L}_n(t) = \sum_{j=1}^{n} \qty(\vec{l}_j + (t_j-t)\frac{\vec{k}_j}{m})\Theta(t_j-t).
    \]
    Similar expressions can be derived for an arbitrary number of $f$-legs.
    Cumulants of the observables defined in~\eqref{eq:momentum_moments} can be derived by taking momentum moments of the phase-space cumulants. In Fourier space, this translates to $\vec{p}\to-\im\nabla_{l}$ and $\int\dd[3]{p}\to\int\dd[3]{l}\dirac(\vec{l})$, i.\,e.~taking the appropriate number of $\vec{l}$-derivatives and setting $\vec{l}=0$.
\end{widetext}

\section{Computing the resummed propagator}
\label{app:macro_details}
In Sec.~\ref{sec:macro_pt}, we introduced the macroscopic formulation of the path-integral approach to classical Hamiltonian $N$-particle systems originally developed in~\cite{Lilow2019}. For completeness, we show how to explicitly compute the resummed propagator in this Appendix, following the original work.

Inserting the ansatz~\eqref{eq:Neumann_series},
\[
    \Delta_{f\B}(X_1,X_2) &= \sum_{n=0}^{\infty} \qty[G_{f\B}^\free]^n (X_1,X_2)\\
    &= \identity(X_1,X_2) + \sum_{n=1}^{\infty} \qty[G_{f\B}^\free]^n (X_1,X_2)\\
    &= \identity(X_1,X_2) + \widetilde{\Delta}_{f\B}(X_1,X_2),
\]
into the definition~\eqref{eq:def_resummed_propagator} of the resummed propagator and multiplying both sides with $(\identity-G_{f\B}^\free)$ yields the self-consistency equation
\[
    \widetilde{\Delta}_{f\B}(X_1,X_2) ={}& G_{f\B}^\free(X_1,X_2) \\
    &+ \int\dd X G_{f\B}^\free(X_1,X)\,\widetilde{\Delta}_{f\B}(X,X_2).
\]
Since the interaction potential and, by extension, the $\B$-field are assumed to be independent of momentum, the internal momentum integration can be carried out easily, replacing $G_{f\B}^\free$ with $G_{\rho\B}^\free$ in the integrand. If, in addition, our system is statistically homogeneous, the spatial integration will be a convolution. In Fourier space $\widetilde{\Delta}_{f\B}$ then has the form
\begin{multline}
    \widetilde{\Delta}_{f\B}(\vec{k}_1,\vec{l}_1,t_1,\vec{k}_2,\vec{l}_2,t_2) \\= (2\pi)^6\dirac(\vec{k}_1+\vec{k}_2)\dirac(\vec{l}_2)\,\widetilde{\Delta}_{f\B}(\vec{k}_1,\vec{l}_1,t_1,t_2).
\end{multline}
We are thus left with the one-dimensional Volterra integral equation
\[
    \widetilde{\Delta}_{\rho\B}(\vec{k},t_1,t_2) ={}& G_{\rho\B}^\free(\vec{k},t_1,t_2) \\
    &+ \int\dd t\,G_{\rho\B}^\free(\vec{k},t_1,t)\,\widetilde{\Delta}_{\rho\B}(\vec{k},t,t_2),
\]
after also integrating out the external momentum.
If the time-dependence of $G_{f\B}^\free$ is only on $\Delta t = t_1-t_2$, this equation again becomes a convolution and can be solved using Laplace transforms \cite{Lilow2019},
\[\label{eq:Laplace_solution}
    \widetilde{\Delta}_{\rho\B}(\vec{k},\Delta t) = \mathcal{L}^{-1}_s\qty[\frac{\mathcal{L}_{\Delta t}[G_{\rho\B}^\free(\vec{k},\Delta t)](s)}{1-\mathcal{L}_{\Delta t}[G_{\rho\B}^\free(\vec{k},\Delta t)](s)}](\Delta t).
\]
In general, however, it needs to be solved numerically by discretizing time and solving the resulting matrix equation.

\section{Interaction potential and time coordinate for cosmic structure formation}
\label{app:time_and_potential}

As in most approaches to cosmic structure formation, be it numerical or analytical, we restrict ourselves to the subhorizon, slow-motion, weak-field regime of general relativity, where perturbative approximations are applicable (see e.\,g.~\cite{Mukhanov1992} for a review). In standard cosmology, the background metric is homogeneous and isotropic, with cosmic structure inducing small perturbations. The latter can be decomposed into scalar, vector and tensor contributions, which evolve independently at linear order in relativistic perturbation theory. In order to describe the system in terms of an interaction potential and make use of the formalism described in section \ref{sec:theory}, it is thus sufficient to restrict to scalar perturbations.
In conformal Newtonian gauge and units where $c=1$, the perturbed line element reads
\[
    \dd s^2 &= g_{\mu\nu}\,\dd x^\mu\,\dd x^\nu\\
    &= a^2\qty(-(1+2\Psi)\,\dd \tau^2 + (1-2\Phi)\,\delta_{ij}\,\dd x^i\,\dd x^j),
\]
with the Kronecker delta $\delta_{ij}$ and using Einstein's summation convention. The scalar perturbations $\Psi$ and $\Phi$ depend on spacetime, and the scale factor $a$ on time. At the background level, Einstein's field equations yield the Friedmann equations governing the evolution of $a$. Restricting ourselves to cold dark matter, i.\,e.~ignoring relativistic matter contributions, the linear order field equations yield $\Psi = \Phi$ and
\[
    \laplace\Phi - 3\,\H\,(\partial_\tau\Phi+\H\Phi) = 4\pi G a^2 \bar\rho_m\delta,
\]
with the mean mass density $\bar{\rho}_m$, the density contrast~$\delta$ and the conformal Hubble parameter $\H = \partial_\tau a/a$. On subhorizon scales, the second term on the left-hand side is typically neglected, resulting in a Poisson-type equation,
\[
    \laplace \Phi = 4\pi G a^2 \bar{\rho}_m\delta.
\]
A point particle moving through spacetime will extremize its proper time. Consequently, its action reads
\[
    S = \int \dd\tau\,a(\tau) \qty[\frac{m}{2}\qty(\pdv{\vec{q}}{\tau})^2 - m\Phi - mc^2]
\]
in the nonrelativistic limit and comoving coordinates~$\vec{q}$ with respect to the background. It can be useful to choose a different time coordinate. In superconformal time \cite{Shandarin1981}, $\dd t_\mathrm{s} = \dd\tau/a$, with the corresponding momenta $\vec{p}_\mathrm{s} = \partial_{t_\mathrm{s}} \vec{q}$, for example, the kinetic term of the Lagrangian has the same form as that of a particle moving in a flat spacetime. The time dependence of the background is then fully absorbed into the potential. Alternatively, one might want to remove the time dependence of the potential altogether. While this is impossible for a general cosmology, it can be achieved in good approximation for late time cosmologies using the well-known time variable \cite{Nusser1998}
\[
    \eta \coloneq \ln \frac{D_+}{D_+^\ini},
\]
where $D_+$ denotes the linear growth factor. Defining the (noncanonical) momentum
\[
    \vec{p} \coloneq m \dv{\vec{q}}{\eta},
\]
the Hamiltonian equations yield \cite{Lilow2019}
\[
    \dv{\vec{p}}{\eta} = -m\nabla\tilde{\Phi} + \qty(1-\frac{3}{2}\frac{\Om}{f_+^2}) \vec{p}
    \approx -m\nabla\tilde{\Phi} - \frac{\vec{p}}{2},
\]
with the matter density parameter $\Om$ and 
\[
    f_+ \coloneq \dv{\ln D_+}{\ln a}.
\]
In the last step, we have used the approximation $\Om/f_+^2 \approx 1$, which has been demonstrated to hold very well for late-time cosmologies \cite{Scoccimarro1998}. Note that this approximation is not necessary, it merely simplifies analytical calculations. The rescaled potential fulfils the Poisson equation
\[
    m\laplace\tilde{\Phi} = \frac{3}{2}\frac{\Om}{f_+^2} \delta \approx \frac{3}{2} \delta.
\]
The solution for a point particle yields the potential energy $v = m\tilde\Phi$ given in~\eqref{eq:point_source_potential}.
Ignoring the potential, the solution \eqref{eq:free_trajectories} of the free equations of motion is replaced by
\[
        \vec{q}^\free(\eta)  = \vec{q}^\ini + \gqp(\eta)\,\vec{p}^\ini \qq{and} \vec{p}^\free(\eta) = \gpp(\eta)\,\vec{p}^\ini,
\]
with the free propagators~\cite{Fabis2015}
\[
    \gqp(\eta_1-\eta_2) &= 2\qty(1-\e^{-\frac{1}{2}(\eta_1-\eta_2)}) \Theta(\eta_1-\eta_2), \\
    \gpp(\eta_1-\eta_2) &= \e^{-\frac{1}{2}(\eta_1-\eta_2)} \Theta(\eta_1-\eta_2).
\]

\section{Initial conditions}
In this Appendix, we give details on the sampling of an initial $N$-particle phase-space density for the two approaches presented in Secs.~\ref{sec:old_ICs} and \ref{sec:new_ICs}, respectively.

\subsection{Gaussian initial conditions in the density and velocity fields}
\label{app:old_ICs}
For the initial conditions presented in~\ref{sec:old_ICs}, the Poisson sampling of an initial $N$-particle phase-space density was presented in~\cite{Fabis2015,Bartelmann2016}. We summarize the key steps in the following.

Consider some realization $\varphi = (\delta^\ini, \vec{v}^\ini)^\transp$ of the density and velocity fields drawn from the ensemble characterized by these statistics. We want to sample particles from the ensemble in such a way that their statistics recover those of $\varphi$ in the limit of large $N$. The probability of finding a particle at $\vec{q}_j$ should hence be proportional to the density~$\rho = \bar{\rho}(1+\delta)$ at that position. For any given $\vec{q}_j$, the associated momentum is given by $m\vec{v}^\ini(\vec{q}_j)$. Therefore, the probability of finding the particle at $\vec{q}_j$ with momentum~$\vec{p}_j$ is
\[
    P(\vec{q}_j,\vec{p}_j \mid \varphi(\vec{q}_j)) = \frac{1}{V} \qty(1+\delta^\ini(\vec{q}_j)) \, \dirac\qty(\vec{p}_j-m\vec{v}^\ini(\vec{q}_j)).
\]
The normalization was chosen by first restricting to a finite volume $V$.
We independently draw $N$ particles in a Poisson-sampling process. Using the notation introduced in~\eqref{eq:bundle_notation}, the conditional probability for the whole ensemble reads
\[\label{eq:P_conditional}
    P(\bm{x}^\ini \mid \bm{\varphi}) = \prod_{j=1}^N P\qty(x_j^\ini \mid \varphi_j=\varphi\qty(\vec{q}_j^\ini)).
\]
In the end, we will take the limit $N, V \to \infty$ at constant $\bar{\rho} = N/V = \bar{\rho}_m/m$.
The initial $N$-particle phase-space probability density $\varrho_N(\bm{x}^\ini) = P(\bm{x}^\ini)$ is thus
\[\label{eq:P_ini}
    \varrho_N(\bm{x}^\ini) &= \int \dd\bm{\varphi}\,P(\bm{x}^\ini \mid \bm{\varphi})\,P(\bm{\varphi})
\]
with 
\[
    P(\bm{\varphi}) = \frac{1}{\sqrt{\det 2\pi \mathbf{C}}} \exp(-\frac{1}{2}\bm{\varphi}^\transp \mathbf{C}^{-1}\,\bm{\varphi}),
\]
where the covariance $\mathbf{C}$ is given by
\[
    \mathbf{C} = \sum_{j,k=1}^{N} \mat{C}\qty(\vec{q}_j^\ini,\vec{q}_k^\ini) \otimes (e_j\otimes e_k).
\]
The integrals in~\eqref{eq:P_ini} can be solved by expressing $P(\bm{\varphi})$ in terms of its Fourier transform, i.\,e.~the characteristic function
\[
    \mathcal{F}[P(\bm{\varphi})](\bm{s}) = \exp(\frac{1}{2}\bm{s}^\transp\mathbf{C}\bm{s}),
\]
resulting in an expression of the form
\begin{multline}
    \varrho_N(\bm{x}^\ini) = \frac{V^{-N}}{\sqrt{\det 2\pi m^2\, \mathbf{C}_{\bm{v\otimes v}}}}\, \mathcal{C}(-\im\bm{\nabla}_{\bm{p}^\ini})\\\times\exp(-\frac{1}{2 m^2} {\bm{p}^\ini}^\transp \mathbf{C}_{\bm{v\otimes v}}^{-1}\,\bm{p}^\ini),
\end{multline}
with some polynomial $\mathcal{C}$. For more details, refer to the Appendix of~\cite{Bartelmann2016}.

\subsection{Gaussian initial conditions in the phase-space density}
\label{app:new_ICs}
Given a Gaussian distribution for the phase-space density $f$, characterized by its cumulants $\mu_{f}$ and $C_{ff}$, the initial conditions of a corresponding particle ensemble can be obtained by Poisson-sampling the phase-space distribution directly,
\[
    P(x_i \mid f^\ini] = \frac{1}{N}\,f^\ini(x_i).
\]
\begin{widetext}
    The resulting $N$-particle phase-space density is
    \[
        \varrho_N(\bm{x}^\ini) &= \int\DD f^\ini\,\prod_{i=1}^{N}\,\bigg[P(x_i \mid f^\ini]\bigg]\,P[f^\ini]\\
        &= \frac{1}{N^N}\sum_{k=0}^{\lfloor N/2 \rfloor} \frac{1}{2^k\,k!\,(N-2k)!} \sum_{\sigma\in S_N} \qty(\prod_{i=1}^{k}C_{ff}(x_{\sigma(2i-1)},x_{\sigma(2i)})) \qty(\prod_{j=2k+1}^{N} \mu_f(x_{\sigma(j)})),
    \]
    by Wick's theorem, where $P[f^\ini]$ is the Gaussian probability functional of $f^\ini$ and $S_N$ denotes the permutation group of $\{1,\dots,N\}$.
    \section{Feynman diagrams for the one-loop spectra}\label{app:self_energies}
    Using the initial conditions defined in~\ref{sec:new_ICs}, there are nine diagrams contributing to the equal-time power spectra,\footnote{In~\cite{Lilow2018,Daus2025}, there are two more diagrams, because of the non-Gaussianities in the initial conditions~\ref{sec:old_ICs}.}
    \[
        \Delta G_{ff}^\text{(1-loop)} ={}&
        \frac{1}{2}\,\begin{tikzpicture}[baseline=-.75mm]
            \draw[thick, {Latex[length=3mm/2,width=3mm/2]}-] (.75/4,0) -- (.5,0);
            \draw[thick] (0,0) -- (.5,0);
            \draw[thick, -{Latex[length=3mm/2,width=3mm/2]}] (1.5,0) -- (1.25/4+1.5,0);
            \draw[thick] (1.5,0) -- (2,0);
            \draw[thick] (1,0) circle (.5);
            \filldraw (.5,0) circle (.05);
            \filldraw (1.5,0) circle (.05);
            \filldraw (1,.5) circle (.05);
            \filldraw (1,-.5) circle (.05);
            \draw[thick, -{Latex[length=3mm/2,width=3mm/2]}] (1-.5/1.414+.001,.5/1.414+.0008)--(1-.5/1.414,.5/1.414);
            \draw[thick, -{Latex[length=3mm/2,width=3mm/2]}] (1+.5/1.414-.001,.5/1.414+.0008)--(1+.5/1.414,.5/1.414);
            \draw[thick, -{Latex[length=3mm/2,width=3mm/2]}] (1-.5/1.414+.001,-.5/1.414-.0008)--(1-.5/1.414,-.5/1.414);
            \draw[thick, -{Latex[length=3mm/2,width=3mm/2]}] (1+.5/1.414-.001,-.5/1.414-.0008)--(1+.5/1.414,-.5/1.414);
        \end{tikzpicture}
        + 2\,\begin{tikzpicture}[baseline=-.75mm]
            \draw[thick, {Latex[length=3mm/2,width=3mm/2]}-] (.75/4,0) -- (.5,0);
            \draw[thick] (0,0) -- (.5,0);
            \draw[thick] (1,0) circle (.5);
            \filldraw (.5,0) circle (.05);
            \filldraw (1.5,0) circle (.05);
            \filldraw (1,.5) circle (.05);
            \draw[thick, -{Latex[length=3mm/2,width=3mm/2]}] (1-.5/1.414+.001,.5/1.414+.0008)--(1-.5/1.414,.5/1.414);
            \draw[thick, -{Latex[length=3mm/2,width=3mm/2]}] (1+.5/1.414-.001,.5/1.414+.0008)--(1+.5/1.414,.5/1.414);
            \draw[thick, -{Latex[length=3mm/2,width=3mm/2]}] (1,-.5)--(1-.001-.1,-.5);
            \draw[thick, {Latex[length=3mm/2,width=3mm/2]}-{Latex[length=3mm/2,width=3mm/2]}] (1.625,0) -- (1.5+1.75/2,0);
            \draw[thick] (1.5,0) -- (2.5,0);
            \filldraw (2,0) circle (.05);
        \end{tikzpicture}
        + \begin{tikzpicture}
            \draw[thick, {Latex[length=3mm/2,width=3mm/2]}-] (.75/2,0) -- (1,0);
            \draw[thick] (0,0) -- (1,0);
            \filldraw (1,0) circle (.05);
            \draw[thick, {Latex[length=3mm/2,width=3mm/2]}-{Latex[length=3mm/2,width=3mm/2]}] (1.25+.25/2,0) -- (1.25+1.75/2,0);
            \draw[thick] (1,0) -- (2.25,0);
            \filldraw (1.75,0) circle (.05);
            \draw[thick] (1,.5) circle (.5);
            \filldraw (1,1) circle (.05);
            \draw[thick, {Latex[length=3mm/2,width=3mm/2]}-] (.5,.449)--(.5,.451);
            \draw[thick, {Latex[length=3mm/2,width=3mm/2]}-] (1.5,.449)--(1.5,.451);
        \end{tikzpicture}
        + 2\,\begin{tikzpicture}[baseline=-.75mm]
            \draw[thick, {Latex[length=3mm/2,width=3mm/2]}-] (.75/4,0) -- (.5,0);
            \draw[thick] (0,0) -- (.5,0);
            \draw[thick] (1,0) circle (.5);
            \filldraw (.5,0) circle (.05);
            \filldraw (1.5,0) circle (.05);
            \filldraw (1,.5) circle (.05);
            \draw[thick, -{Latex[length=3mm/2,width=3mm/2]}] (1-.5/1.414+.001,.5/1.414+.0008)--(1-.5/1.414,.5/1.414);
            \draw[thick, -{Latex[length=3mm/2,width=3mm/2]}] (1+.5/1.414-.001,.5/1.414+.0008)--(1+.5/1.414,.5/1.414);
            \draw[thick, -{Latex[length=3mm/2,width=3mm/2]}] (1,-.5)--(1-.001-.1,-.5);
            \draw[thick, -{Latex[length=3mm/2,width=3mm/2]}] (1.5,0) -- (1.25/4+1.5,0);
            \draw[thick] (1.5,0) -- (2,0);
        \end{tikzpicture}
        + \frac{1}{2}\,\begin{tikzpicture}
            \draw[thick, {Latex[length=3mm/2,width=3mm/2]}-] (.75/2,0) -- (1,0);
            \draw[thick] (0,0) -- (1,0);
            \filldraw (1,0) circle (.05);
            \draw[thick, {Latex[length=3mm/2,width=3mm/2]}-] (2-.75/2,0) -- (1,0);
            \draw[thick] (1,0) -- (2,0);
            \draw[thick] (1,.5) circle (.5);
            \filldraw (1,1) circle (.05);
            \draw[thick, {Latex[length=3mm/2,width=3mm/2]}-] (.5,.449)--(.5,.451);
            \draw[thick, {Latex[length=3mm/2,width=3mm/2]}-] (1.5,.449)--(1.5,.451);
        \end{tikzpicture} \\
        &+ \begin{tikzpicture}[baseline=-.75mm]
            \draw[thick, {Latex[length=3mm/2,width=3mm/2]}-] (.75/4,0) -- (.5,0);
            \draw[thick] (0,0) -- (.5,0);
            \draw[thick] (1,0) circle (.5);
            \filldraw (.5,0) circle (.05);
            \filldraw (1.5,0) circle (.05);
            \draw[thick, -{Latex[length=3mm/2,width=3mm/2]}] (1,.5)--(1-.001-.1,.5);
            \draw[thick, -{Latex[length=3mm/2,width=3mm/2]}] (1,-.5)--(1-.001-.1,-.5);
            \draw[thick, {Latex[length=3mm/2,width=3mm/2]}-{Latex[length=3mm/2,width=3mm/2]}] (1.625,0) -- (1.5+1.75/2,0);
            \draw[thick] (1.5,0) -- (2.5,0);
            \filldraw (2,0) circle (.05);
        \end{tikzpicture}
        + 2\,\begin{tikzpicture}[baseline=-.75mm]
            \draw[thick, {Latex[length=3mm/2,width=3mm/2]}-] (.75/4,0) -- (.5,0);
            \draw[thick] (0,0) -- (.5,0);
            \draw[thick] (1,0) circle (.5);
            \filldraw (.5,0) circle (.05);
            \filldraw (1.5,0) circle (.05);
            \draw[thick, -{Latex[length=3mm/2,width=3mm/2]}] (1,.5)--(1+.101,.5);
            \draw[thick, -{Latex[length=3mm/2,width=3mm/2]}] (1,-.5)--(1-.001-.1,-.5);
            \draw[thick, {Latex[length=3mm/2,width=3mm/2]}-{Latex[length=3mm/2,width=3mm/2]}] (1.625,0) -- (1.5+1.75/2,0);
            \draw[thick] (1.5,0) -- (2.5,0);
            \filldraw (2,0) circle (.05);
        \end{tikzpicture}
        + 2\,\begin{tikzpicture}
            \draw[thick, {Latex[length=3mm/2,width=3mm/2]}-] (.75/2,0) -- (1,0);
            \draw[thick] (0,0) -- (1,0);
            \filldraw (1,0) circle (.05);
            \draw[thick, {Latex[length=3mm/2,width=3mm/2]}-{Latex[length=3mm/2,width=3mm/2]}] (1.25+.25/2,0) -- (1.25+1.75/2,0);
            \draw[thick] (1,0) -- (2.25,0);
            \filldraw (1.75,0) circle (.05);
            \draw[thick] (1,.5) circle (.5);
            \draw[thick, -{Latex[length=3mm/2,width=3mm/2]}] (1, 1) -- (1-.101,1);
        \end{tikzpicture}
        + \begin{tikzpicture}[baseline=-.75mm]
            \draw[thick, {Latex[length=3mm/2,width=3mm/2]}-] (.75/4,0) -- (.5,0);
            \draw[thick] (0,0) -- (.5,0);
            \draw[thick] (1,0) circle (.5);
            \filldraw (.5,0) circle (.05);
            \filldraw (1.5,0) circle (.05);
            \draw[thick, -{Latex[length=3mm/2,width=3mm/2]}] (1,.5)--(1+.101,.5);
            \draw[thick, -{Latex[length=3mm/2,width=3mm/2]}] (1,-.5)--(1-.001-.1,-.5);
            \draw[thick, -{Latex[length=3mm/2,width=3mm/2]}] (1.5,0) -- (1.25/4+1.5,0);
            \draw[thick] (1.5,0) -- (2,0);
        \end{tikzpicture}
        .
    \]
    All of these diagrams contain multiple terms, which grow at a different rate with time. If one is interested in late-time statistics, only the fastest growing modes have to be taken into account~\cite{Bernardeau2002}. The first three diagrams are the only ones which contain terms of order $D_+^4$. In~\ref{sec:one_loop}, we restrict to these terms.
\end{widetext}

\bibliography{references_bibtex.bib}

\end{document}